# The Biosynthetic Order of Amino Addition to the Genetic Code

Brian K. Davis

Research Foundation of Southern California Inc.
5580 La Jolla Boulevard
La Jolla, CA 92037, U.S.A.

1 Table
1 Figure
22 pages

Abbreviations: A, adenine; C, cytosine; G, guanine; U, uridine; N, any standard nucleotide; R, purine; Y, pyrimidine; AICARP, aminoimidazole carboxamide ribonucleotide; Ala, alanine; Arg, arginine; Asn, asparagine; Asp, aspartate; Cys, cysteine; f-Met, N-formyl-methionine; Glu, glutamate; Gln, glutamine; His, histidine; Ile, isoleucine; Leu, leucine; Lys, lysine; Met, methionine; Phe, phenylalanine; Pro, proline; Sec, selenocysteine; Ser, serine; Thr, threonine; Trp, tryptophan; Tyr, tyrosine; Val, valine; C, coefficient of variation; 8, mean of difference; P, probability; tRNA-I, type I; -II, type II; -A, core group A, -B, group B, -C, group C; -D, group D. Superscripts on amino acids, number of reaction steps in biosynthesis, which corresponds generally to stage of addition to the code.

A previously formulated model for the evolution of the genetic code was shown to clarify why base triplets for some precursor amino acids differ by a single base from product amino acid codons, while others show no homology. First, the model indicated that the direction of code evolution changed on expansion from the N-fixers code (stage 2). Growth of the code from 16 codons in the NAN column (N, any standard nucleotide) proceeded by assignment of codons in the GNN, ANN, CNN and UNN rows. Expansion phase (stage 4 to 7) precursor/product pairs that spanned this shift included aspartate/threonine, aspartate/ methionine, and glutamate/proline. Both 5'- and mid-base differ in the codons of each pair. Second. post-expansion additions (stage 9 to 14) required codon reassignment, eliminating initial correlations. Codons for the post-expansion pair, aspartate (glutamate)/arginine, differ at both 5'- and mid-base sites. Third, the distribution of core structure groups among acceptors indicated that variant tRNA specific for a sibling, rather than precursor, commonly participated as cofactors in early amino acid synthesis. Sibling pairs, rather than precursor/ product pairs, then exhibit codon correlations. On removing these sources of variation, highly significant correlations emerged between codons assigned to biosynthetically related amino acids, most conspicuously in expansion phase.

## 1. Introduction

All twenty-two kinds of amino acids incorporated into protein are synthesized on pathways that begin in central metabolism. The distribution of their sources within central metabolism is also suggestive, with respect to how the first proteins and genetic code formed. Almost two-thirds (14/22) of encoded amino acids arise from the citrate cycle. Branch pathways extending only 1 or 2 reaction steps from the citrate cycle produce the four significant N-fixing amino acids, Asp$^1$, Asn$^2$, Glu$^1$, Gln$^2$, plus Ala$^1$. Consistent with the antiquity of the N-fixers, they are precursors to nearly half (9/20) the amino acids in the code. The 'before and after' order of codon sets assigned to 11 distinct precursor/product amino acid pairs, revealed (Davis, 1999) that 5 transitions emanated from the NAN codon set, with no back-transitions. This identified these triplets as the earliest set of codons within the code ($P = 7.8 \times 10^{-3}$). Significantly, all four N-fixers were encoded by this set. The first code was then inferred to consist of the 16 codons in the NAN set and 4 N-fixing amino acids, plus a TER signal. Expansion to all 64 codons was deduced to occur by addition of increasingly hydrophobic amino acids, synthesized on progressively longer pathways. Amino acids with basic groups and ring structures in their side chain are synthesized on the longest pathways, and they were judged to be post-expansion additions to the code.

Nirenberg et al. (1966) noted that some biosynthetically related amino acids had similar codons in the standard code. Dillon (1973) and Wong (1975) interpreted this to imply that the code had coevolved with amino acid pathways. In particular, Wong attributed codon correlation to acceptor molecules participating as cofactors in early amino acid synthesis. Previously, Wilcox & Nirenberg (1968) had established that $tRNA^{Gln}$ served as a cofactor in Gln synthesis in three Bacteria species. This mechanism is now known to extend to other amino acids, notably Asn, in both Bacteria and Archaea (Schon et al, 1988; Guillon et al., 1992; Curnow et al., 1996; Gagnon et al., 1996; Wilting et al., 1997). tRNA also serves as a cofactor for amino acids in reactions unrelated to amino acid synthesis and even in reactions not involving amino acids (Danchin, 1989).

Wong (1975) demonstrated a statistically significant elevation in codon homology between codons of biosynthetically related amino acids ($P < 2 \times 10^4$). However, this was achieved by excluding 5 of the 13 precursor/product pairs identified. A recent reappraisal of codon correlations in a complete, partly modified set of precursor/product pairs, adjusted for code degeneracy, revealed they were non-significant, with P = 0.12 to 0.62 (Amirnovin, 1997; Di Giulio & Medugno, 2000; Ronnenberg et al., 2000). This led Ronnenberg and coworkers to question whether the biosynthetic theory of the genetic code was "fact or artifact?"

These findings prompted the present evaluation of the biosynthetic model. Possibly, the model would clarify why some precursor/product pairs had similar codons, while others do not. Two mechanisms arose that could separate codons assigned to some precursor/product amino acid pairs. First, expansion from the 16 codon N-fixers code resulted in a change in the direction of code evolution. Codons for any precursor/product pair that spanned this change would be expected to differ by more than one base. Second, post-expansion additions to the code occurred by capture of assigned codons. Existing correlations with codons reassigned in this process would be extinguished, without new ones necessarily replacing them. In the event that precursor/product pairs excluded by Wong (1975) were produced by these means, elevated homology between codons for precursor/product pairs would be established as a feature of the genetic code,

Codon assignment patterns within the code will be shown to retain strong evidence of biosynthetic kinship between amino acids. To avoid phase-specific effects related to the direction of code evolution and mechanism of codon assignment, each phase was examined separately in this paper. The distribution of tRNA core structure groups, identified in archaeal thermophiles and related species (Saks & Sampson, 1995), among families of related amino acids was also considered, and provided further evidence of tRNA cofactors in early pathways for amino acid synthesis.

## 2. Phases of code evolution

Figure 1 shows four phases identified in the evolution of the genetic code (Davis, 1999). In the N-fixers code, base triplets of the NAN set were assigned to Asp$^1$, Asn$^2$, Glu$^1$ and Gln$^2$. It points to occurrence of an earlier, pre-code phase of protein synthesis involving translation on a poly A template by a 'universal' acceptor, $tRNA^{(Asp,Asn,Glu,Gln)}_{UUU}$, specific for the codon, AAA. Protein synthesis at this stage preceded gene replication and formation of specific 'synthetases'. A single tRNA species was apparently charged with Asp$^1$, Asn$^2$, Glu$^1$ or Gln$^2$.

FIGURE 1

Three of four acceptors had a group D core, at formation of the N-fixers code, indicating the ancestral, 'universal' acceptor also had a group D core. Pre-code translation on a poly A template would result in a random sequence polypeptide chain. Being polyanionic poly (Asp,Asn,Glu,Gln) could anchor uncharged amide residues to a positively charged mineral surface, in the pre-cell surface system visualized by Wachterhauser (1988, 1992, 1997, 1998).

Increased efficiency is gained in this early N-fixation mechanism on restricting the mole ratio of carboxyl- and amide-residues in poly (Asp,Asn,Glu,Gln) to one that jointly optimizes charge attraction to a mineral surface and amide residue content (Davis, 1999). This necessitates template-directed synthesis of ordered peptide sequences and so provides a molecular basis for devolpment of the N-fixers code. A notable feature of this code is the presence of a single group A core, in tRNA-$A^{Asn}_{UUU}$ (Fig. 1). As tRNA-$A^{Asn}_{UUU}$ shares anticodon, UUU, with the ancestral acceptor, tRNA-$D^{(Asp,Asn,Glu,Gln)}_{UUU}$, its unique core structure would provide an identity element allowing Asn-acceptor recognition by Asn 'synthetase'. Synthesis of uncharged (non-surface binding) amide residues in the primordial surface system seemed to follow misacylation of tRNA-$A^{Asn}_{UUU}$ and tRNA-$D^{Gln}_{3'GUU}$ by $Asp^1$ and $Glu^1$, respectively, and amidation at their terminal carboxyl.

The small N-fixers code shows refinements that appear to result from selection forces favouring charge confinement. Both carboxyl-residues have codons in the GAN set. Any 5'-anticodon base substitution is then neutral. Misreading of the 5'-base during translation would also substitute one acidic residue for another. This codon assignment pattern produces a polarity difference between the GAN set, and AAN and CAN sets, akin to the conspicuous polarity/hydrophobicity gradient in the standard code. The mean free energy change, $\Delta F_T$, for transfer of Asp and Glu from an aqueous to non-polar solvent (dielectric constant, 2.0) is 8.7 kcal/mol, while the amides have a $\Delta F_T$ of 4.5 kcal/mol (Tolstrup et al. 1994). Assignment of weak-binding codons, UAN, to Ter, which are read infrequently, parallels standard codon assignments. Initial limitation to codons of the NAN set, in the N-fixers code, contained an in-built error avoidance feature. With only 16 of 64 base triplets assigned, at this stage of code evolution, it follows that three-quarters of base substitutions among randomly assigned codons would yield an unreadable triplet. Missense codons of this sort arrest template translation (Bretscher et al., 1965). In the N-fixers code, only one-third of base substitutions (any mid-base change) yields an unreadable triplet. The mid-base of each codon is an A, in this early code (Fig. 1). This is compatible with the N-fixers code being commaless, at some stage.

Consistent with this, ancestral ferredoxin seems to have had repetitious residue sequence (Eck & Dayhoff, 1966). In a commaless code, translation is phased locally. Thus, synthesis could restart, following an interruption, at any downstream codon and still produce a functional protein. Translation conceivably progressed to 5' initiation in the synthesis of N-fixer peptides. An attachment site could then be positioned within the peptide chain. 5'-Initiation represented a precondition for assignment of codons not restricted to a mid-A and was consequently essential for expansion from the N-fixers code.

Expansion from a code of 16 to all 64 possible codons was associated with addition of amino acids synthesized on pathways extending 4 to 7 reaction steps from the citrate cycle (Davis, 1999). The sole exception being $Ala^{4.1}$, which forms upon amination of pyruvate in standard prokaryote pathways (Greenberg, 1969). There are three expansion phase amino acids in the Asp family; $Thr^{6}$, $Ile^{7}$, $Met^{7}$ (Fig, 1). All have codons on the ANN row of the code, similar to $Asn^{2}$. Also, like Asn, each has a group A acceptor. Occurrence of related codons and acceptors with homologous core structure among Asp family siblings provides compelling evidence for participation of a variant $tRNA\text{-}A^{Asn}$ as a cofactor in early amino acid synthesis. Significantly, this mechanism interlinks codon assignment patterns, synthesis of new amino acids and diversification of tRNA specificity.

Each variant $tRNA\text{-}A^{Asn}$ was seemingly misacylated with $Asp^{1}$. Synthesis of $Thr^{6}$, $Ile^{7}$ and $Met^{7}$ then proceeded with a tRNA cofactor, as in $Asn^{2}$ synthesis (Schon et al, 1988; Curnow et al., 1989; Gagnon et al., 1996). $Alanine^{4.1}$ and $Val^{4}$ also use group A acceptors (Saks & Sampson, 1995). Both originate from pyruvate in extant pathways. However, $Ala^{4.1}$ also forms by decarboxylation of $Asp^{1}$ (Doctor & Oro, 1972). Formation of $Val^{4}$ from $Asp^{1}$ by this mechanism is also conceivable (§4). Use of group A acceptors by $Ala^{4.1}$ and $Val^{4}$, therefore, lends credence to the antiquity of the Doctor-Oro pathway, and inclusion of both in the original Asp family. $Proline^{4}$ has a group D acceptor, similar to both its precursor, $Glu^{1}$, and sibling, $Gln^{2}$. As with expansion phase members of the Asp family, synthesis of $Pro^{4}$ appeared to occur by misacylation of a variant acceptor for the amide with its acidic precursor. Close sequence alignment between tRNA-D$^{Pro}$ and tRNA-D$^{Gln}$ supports this (Saks & Sampson, 1995; Saks et al., 1998).

Glycine$^{5}$ is a stage 5 addition to the code. Its acceptor, tRNA-C$^{Gly}_{3'CCU}$, has a group C core. This places formation of a group C core structure in expansion phase (Fig. 1). Removal of a hydroxy-methyl group from Ser$^{4}$ is the standard route to Gly$^{5}$. tRNA$^{Ser}$ is a type II tRNA. This excludes it as an antecedent of tRNA-C$^{Gly}$ and, therfore, as a cofactor in its synthesis. A Val$^{4}$ isoacceptor has a group C core in thermophilic Archaea (Saks & Sampson, 1995). Its anticodon, 3'-CAG, restricts this acceptor to the codon pair, GUY, among Gly$^{5}$ codons. Hence, it appeared to follow entry of this amino acid into the code. This implies tRNA-A$^{Val}_{3'CAU}$ was ancestral among Val$^{4}$ acceptors. Since its 5'-anticodon U could wobble pair with any 3'-codon base, in a simple protein synthesizing system ( Barrell et al., 1980; Andachi et al., 1987), a single acceptor could initially translate all Val$^{4}$ codons, GUN. Cysteine$^{5}$ is another expansion phase product of Ser$^{4}$. The Cys$^{5}$ acceptor core group remains to be determined, however. Comparable to Ser$^{4}$ isoacceptors, Leu$^{7}$ are type II tRNA molecules. tRNA-II$^{Leu}$ also translates triplets in two codon sets (CUN, UUN). No direct link exists between the type II acceptors for Leu$^{7}$ and type I acceptors for Ala$^{4}$ and Val$^{4}$, apart from contiguity between Val/Leu anticodons.

The diaminopimelate pathway leading to Lys$^{10}$ extends 10 reaction steps from the citrate cycle. Consistent with once being an attachment site for a tRNA cofactor, the $\alpha$-carboxyl of Asp$^{1}$ does not react over the length of this path. Sequence homology between tRNA-A$^{Lys}_{3'UAC}$ and tRNA-A$^{Ile}_{UUU}$ (Saks & Sampson, 1995) strengthens this intrepretation. Capture of doublet AAR in the AAN set of 'weak' codons assigned to Asn$^{2}$ would have been facilitated by the greater reactivity of peptidyl transferase with this basic amino acid (Krayevsky & Kukhanova, 1979; Remme & Villems,1985; Davis, 1999 §2.4.2).

From comparative path lengths for synthesis, codons were assigned to a more basic (pK 12.5) amino acid, Arg$^{9}$, before Lys$^{10}$. Arginine$^{9}$ has a group A acceptor (Fig. 1), even though 5 C and 1 N of its atoms are contributed by Glu, which has a group D acceptor. Aspartate$^{1}$ notably contributes 1 N atom at step 8. Thus, Arg$^{9}$ acceptors show homology with acceptors for Asp family siblings, lending qualified support for 'the closest direct antecedent' rule as the basis for codon correlations (Ronnenberg et al., 2000).

Both aromatic amino acids, Phe$^{11}$ and Tyr$^{11}$, have group B acceptors . Some of their isoacceptors also exhibit close sequence homology, suggesting they had a common ancestor (Saks & Sampson, 1995). The phosphoenolpyruvate (central trunk) carboxyl remains unmodified over the length of the shikimate pathway, consistent with once being a linkage site for a tRNA cofactor. Capture of related 'weak' codon pairs, UUY and UAY, by Phe and Tyr (Fig. 1) would have been aided by elevated transferase reactivity with aromatic amino acids (Remme & Villems,1985). Error minimization and selection for incorporation of aromatic amino acids into post-expansion phase proteins also seemed tofavour these assignments.

Consistent with being late entrants to the code, His$^{13}$ and Trp$^{14}$ form on pathways extending 13 and 14 steps from the citrate cycle, respectively. As would be anticipated for latecomers to the code, they have only 1 or 2 codons each in the standard code. No protected carboxyl group occurs in the His pathway. However, Gln reacts with phosphoribulosyl formimino AICAR (aminoimidazole carboxamide ribonucleotide) at step 7 (fifth reaction step from ribose-5-phosphate, pentose cycle). Retention of a variant tRNA-D$^{Gln}$ cofactor, with its transfer to His $\alpha$-carboxyl, formed at step 12, provides a mechanism for His acquiring Gln codons CAY and a group D acceptor. Trptophan$^{14}$ synthesis branches from the shikimate pathway at chorismic acid (Greenberg, 1969). The codon pair UGR was assigned to Trp at stage-14 (Fig. 1). They differ by a single substitution from Ser$^{4}$ codons in the UCN set (by two substitutions from the codon pair, AGY), consistent with Ser$^{4}$ being its 'closest direct antecedent'. On the other hand, tRNA trees constructed by Bermudez et al. (1999) and Chaley, et al. (1999) show tRNA-II$^{Ser}$ and tRNA$^{Trp}$ do not have closely related sequences. It appears unlikely, in view of this, that a variant of extant Ser acceptor participated in Trp synthesis.

## 3. Biosynthetic and physical chemical patterns in the code

### (a) ***N-fixers code***

Two precursor/product amino acid pairs occur in the N-fixers code. Codons for the Glu/Gln pair, GAN and CAN (Fig. 1), differ by a 5'-G:C substitution. Aspartate/Asn codons GAY and AAN differ by a 5'-G:A transition plus a 3'-Y:N substitution (a G34U tRNA substitution). The aggregate distance (number of substitutions) between both pairs of anticodons equals three substitutions, giving a mean difference of 1.5 ± 0.47 ($\delta_o$ ± C, where C is the coefficient of variation) per pair (Table 1). On permuting 4 acceptors between these four amino acids, the N-fixers code is found to be one of 24, or 4!, possible codes. All have an aggregate difference between anticodons of 3, or 2, substitutions. Every code thus has a mean difference of 1.5, or less, per precursor/product pair. Among the population of possible codes, all codes have anticodons at least as strongly clustered ($\delta \leq \delta_o$) as the N-fixers code. Hence, it is certain (P = 1) any randomly chosen code will achieve, or exceed, the level of biosynthetic kinship in the N-fixers code. No evidence of anticodon (codon) clustering exists, therefore, among biosynthetically related amino acids in the N-fixers code. Stronger clusters of precursor/product amino acid pairs result when one pair has anticodons 3'-CUU/3'-CUG (codons, GAN/GAY), and the other UUU/3'-GUU (codons, AAN/ CAN). Eight codes ($2^3$) exhibit stronger clusters (P = 0.33), with a smaller mean difference ($\delta$ = 1 < 1.5) between anticodons.

TABLE 1

Cluster formation based on two tRNA identity elements was examined. This increases the number of possible identity element combinations (expanded codes). Clusters with a lower probability of occurrence then become detectable. There are, as noted, 3 group D and 1 group A acceptors in the N-fixers code (Fig. 1). Hence, there are 4 ways (4!/3!1!) of distributing the core groups to four acceptors, each specific for one of four N-fixing amino acids. Joining the number of combinations for acceptor core groups and anticodons yields a space of 96, or 24 x 4, possible expanded codes (Table 1). All expanded codes, the N-fixers included, have precursor/product pairs with DD and DA pairs of core groups. The aggregate difference between Glu/Gln and Asp/Asn pairs thus increases from 3 to 4 differences, giving a mean of $2.0 \pm 0.71$ (Table 1). No evidence of cluster formation based on biosynthetic kinship results. All codes yield a mean difference of 2.0, or less, per pair ($P = 1$). As before, there is a probability of 0.33 (32/96) of a code with a smaller mean difference (8 = $1.5 < 2.0$) than in the N-fixers expanded code.

Charged (Asp, Glu) and non-charged (Asn, Gln) amino acids are clustered more strongly than biosynthetic pairs in the N-fixers code (Table 1). The mean difference between their anticodons is 1.0. Distributing 4 acceptors among 2 charged and 2 non-charged amino acids results in 6, or 4!/2!2!, distinguishable codes. The N-fixers code and one other have a mean of 1.0 difference per pair. In the remaining 4 codes, the mean difference was 1.5.

Wobble pair patterns allow a distinction to be made between the two superior codes. One assigns acceptors with anticodons UUU and 3'-GUU, specific for codon sets AAN and CAN, respectively, to the two charged amino acids. The second uses acceptors with anticodons 3'-CUU and 3'-CUG, specific for codons in the *same* codon set, GAN, for charged amino acids. In the event, Asp$^{1}$ and Glu$^{1}$ acceptors both translate the codon pair GAY, a zero effective mean difference exists between them. The N-fixers code is then optimal for charge confinement, among 6 possible codes ($P = 0.17$).

Confinement of both charged amino acids to the same codon set gives rise to a polarity gradient within the N-fixers code. Aspartate$^{1}$ and Glu$^{1}$, have $\Delta F_T$ values of 9.2 and 8.2 kcal/mol, respectively (Tolstrup et al., 1994), and rank 1 and 2 in polarity among N-fixing amino acids. A G34U substitution separates their anticodons. The low polarity pair Asn$^{2}$ and Gln$^{2}$ rank 3 and 4, with $\Delta F_T$ values of 4.8 and 4.2 kcal/mol, respectively. Their anticodons differ by a U36G substitution. A mean increment in $\Delta F_T$ rank of 1.0 occurs per pair. Permuting acceptors between the four N-fixing amino acids results in possible 24 codes. Eight have a mean increment in rank $\Delta F_T$ of 1.0 per pair. The remaining 16 have a mean difference of 2.0 ranks per pair. Clusters of high and low polarity amino acids equal, or stronger, to that in the N-fixers code arise in 33 per cent (8/24) of all codes (Table 1). In summary, the strongest evidence of clustering within the N-fixers code is provided by charge confinement.

(b) ***Expansion phase***

During expansion from the N-fixers code, increasingly hydrophobic amino acids were assigned codons of the NCN, NGN and NUN sets, broadly in order of assignment (Fig. 1). On ranking these 10 amino acids according to their $\Delta F_T$, the mean increment between successive amino acids within each codon set was only 1.2 ± 0.68 ($8_o \pm C$). This code (stage 7) almost minimized the mean within-set difference in $\Delta F_T$ rank over these three sets of codons. Interchanging $Ala^{4.1}$ and $Gly^{5}$ with $Thr^{6}$ and $Ser^{4}$, between NCN and NGN sets, would minimize the mean within-set increment in rank at 0.78 ± 0.57. Only $3.89 \times 10^4$ of $1.20 \times 10^8$ possible codes have a mean difference between ranked amino acids equal to, or less, than in the stage 7 code. The number of possible codes corresponds here to the number of permissible ways of distributing 12 sets of acceptors (codon sets) to 10 amino acids, which is 12!/2! [2]. To obtain the number of codes with equal, or stronger, clusters than the stage 7 code, codon sets assigned to each amino acid were permuted among amino acids assigned codons within the NCN, NGN and NUN sets. From the arrangement of amino acids in the stage 7 code (Fig. 1), there are 8,640, or 4!3!(5!/2!), equivalent codes. On accumulating the number of codes corresponding to each single interchange ( Ser ↔ Ala, Ser ↔ Gly, Thr ↔ Ala, Thr ↔ Gly) between NCN and NGN sets and one double interchange (Ser,Thr ↔ Ala,Gly) the number of codes with equal, or stronger, clusters than the stage 7 code is obtained. Amino acid addition to the code during expansion phase was thus found to produce a highly significant hydrophobicity gradient over the NCN, NGN and NUN codon sets, $P = 3.25 \times 10^{-4}$ (Table 1).

No charged amino acids were added to the code during expansion phase, spanning stages 4 to 7. The hydrophobicity gradient between codon sets NCN, NGN and NUN, can be attributed to a change in selection forces acting on the code in this interval. Addition of increasingly hydrophobic amino acids to the code would be favour synthesis of globular proteins, containing a hydrophobic interior, and membrane proteins.

Highly significant biosynthetic kinship patterns also arose among expansion phase additions to the code. Seven of 10 amino acids encoded during expansion originate from citrate cycle components. Threonine$^{6}$, Ile$^{7}$ and Met$^{7}$ originate at oxaloacetate. They share the ANN row with Asn$^{2}$, a sibling from the N-fixers code (Fig. 1). α-Ketoglutarate led to Pro$^{4}$, which shares the CNN row with a sibling, Gln$^{2}$, from the N-fixers code. Pyruvate resulted in Ala$^{4.1}$ and Val$^{4}$, both on the GNN row. In addition, Leu$^{7}$ shared the NUN column with its sibling Val $^{4}$. 3-Phospho-D-glycerate (central trunk) gave rise to Ser$^{4}$, Gly$^{5}$ and Cys$^{5}$, and they share the NGN column. Serine$^{4}$ and Cys$^{5}$ also share the UNN row. The distribution of biosynthetically related amino acids in the stage 7 code is strongly clustered. Only 9 base substitutions separate the codon sets for all ten expansion phase amino acids and two N-fixer siblings.

Interchanging rows GNN and UNN of the expansion phase set, NCN, NGN, NUN, produces an equivalent code, with all family relations between amino acids at stage 7 (Fig. 1) preserved. Rows ANN and CNN do not interchange, without severing links between Asn and Gln, in the NAN column, and expansion phase siblings. Equivalent codes result on interchanging columns NCN, NGN and NUN. The location of Val$^{4}$ is fixed by its links with Ala$^{4.1}$ (codon and tRNA core structure homology) and Leu$^{7}$ (codon homology). Alanine$^{4.1}$ and Leu$^{7}$ are freely interchangeable, however. Threonine$^{6}$, Ile$^{7}$ and Met$^{7}$ can be permuted within the ANN row. Serine$^{4}$, Gly$^{5}$ and Cys$^{5}$ can be permutated, excluding two codes that orphan Gly$^{5}$ and Cys$^{5}$ on UCN triplets. The position of Pro$^{4}$ is fixed in row CNN, because Leu$^{7}$ at CUN cannot be switched without severing its link with sibling Val$^{4}$. Combining all permissible variations reveals there are 2,160 codes, or $2 \times 3! \times \frac{3!}{2} \times 3! \times (\frac{4!}{2} - 2)$, with a mean difference of 1.0 base substitution between distinct codon sets (minimum difference) for biosynthetically related amino acids. As noted, there are $1.20 \times 10^{8}$ possible expansion phase codes. Occurrence of a code in which there is a mean difference of only 1.0 base substitution between codons for related amino acid therefore has a probability of $1.80 \times 10^{-5}$ (Table 1).

Acceptor core groups, identified by Saks & Sampson (1995), are known for 6 of 9 pairs of biosynthetically related amino acids in the stage 7 code. Serine$^{4}$ and Leu$^{7}$ acceptors are type II tRNA, whose core group was not specified. Each of the six pairs with known core groups was homologous. With 5 group A and 1 group D acceptors (Fig. 1), homology between all pairs represented 1 of 6, or 6!/5!, possible arrangements. The number of 'codes' based on two tRNA identity elements (anticodon, core) with a mean of 1.0 difference between base triplets for related amino acids remained 2,160. Whereas, the number of possible codes increased to 7.19 x $10^{8}$, or 6 x 1.20 x $10^{8}$. The probability of a code that clustered biosynthetic kinship and tRNA core groups as strongly, or better, than the stage 7 code was, therefore, 3.01 x $10^{-6}$ (Table 1).

(c) ***Codon capture phase***

Codons assigned to post-expansion (stage 9 to 14) additions to the code show strong biosynthetic kinship. Isoleucine/Lys, Gln/His and Phe/Tyr have a common source in central metabolism, acceptors with core group and sequence homology, and codon sets that differ by a single base substitution. Aspartate$^{1}$ and Ser$^{4}$ are the closest direct antecedent of Arg$^{9}$ and Trp$^{14}$, respectively. Acceptor sequence and core group homology was not available for the Ser/Trp pair. Also, the Arg$^{9}$ CGN codon set differed by two base substitutions from members of the Asp family. The mean difference between codon sets for related amino acids was found to be 1.20 $\pm$ 0.35 (Table 1). Equivalent codes include those with Phe and Tyr, Lys and His, and Arg (CGN) and Trp acceptors interchanged. On taking into account the number of codes obtained after permuting all 4 rows and all 4 columns, which preserve pre-existing relationships, the number of equivalent codes was found to be 4,608, or 4! X 4! X $2^{3}$. The number of possible codes formed on independently permuting all rows and all columns, and all seven acceptor specific for the six post-expansion amino acids was 1.45 x $10^{6}$, or 4! x 4! x $\frac{7!}{2}$. There results a probability of 3.17 x $10^{-3}$ that the correlations at stage 14 arose by chance.

A lower probability is achieved on including acceptor core groups. With 3 A, 2 B and 1 D acceptor groups, the number of possible expanded codes was 8.71 x $10^{7}$, or 1.45 x $10^{6}$ x 6!/(3!2!). The number of equivalent codes, which preserve, or strengthen, biosynthetic kinship and tRNA core group correlations at stage 14 remained 4.61 x $10^3$. A probability of 5.29 x $10^{-5}$, or 4.61 x $10^3$/8.71 x$10^7$, exists that anticodon/core group correlations among post-expansion addtions of biosynthetically related amino acids in the stage 14 code arose by chance.

Codons for the basic amino acids $Arg^9$, $Lys^{10}$ and $His^{13}$ occupy four corners of a square in the stage 14 code (Fig. 1). They differ by a mean of 1.7 ± 0.49 base substitutions; each side of the square represents a single base difference and both diagonals a double base difference between codon sets. As all three amino acids are positively charged in a near neutral (pH 6.4) medium, the possibility arose that charge confinement underlay formation of this cluster. There are 22 acceptors with distinct coding specificity (1codon set/acceptor, discounting isoacceptors) in the stage 14 code (Fig. 1). On preserving the set, GAN, assigned previously to both negatively charged amino acids, we find there are 5,985, or 21!/(21-4)!4!, ways of distributing 4 acceptors for charged residues, and 17 for uncharged residues. Individual amino acid identities are suppressed here.

A row, or column, of a set of four amino acids, has codon sets that differ at one site only (given 3'-base substitutions are silent). The set of transitions between each amino acid then has a mean of only 1.0 base difference per pair. In particular, any 5'-base substitution in a column of codons for charged amino acids produces no effective change, while any mid-base substitution results in a change. For a row in the code, these outcomes switch at each site. In a square cluster, by comparison, one substitution at each site will be silent. This implies two substitutions at each site will alter a protein charge distribution. With three base substitutions possible at each site, row and column (3/6 silent) clusters are superior to a square (2/6 silent) in suppressing errors. While a T-, or L-shaped cluster (2/6 silent, overall) is as effective as a square. Given the 3'-base degeneracy pattern at stage 14, the code admits 5 row (GAN and UAR excluded), 36 column, 75 square, 280 tee, and 408 elle configurations of four charged amino acids. A total of 804 equivalent codes occur. Consequently, there is a probability of 0.13, or 804/6.0 x $10^3$, that a distribution of positively charged amino acids equally, or more strongly, clustered could form randomly.

When $\Delta F_T$ values were used to rank post-expansion amino acids, a gradient of increasing polarity/hydrophobicity arose between ANN, CNN and UNN codon sets. It was only a single exchange (Arg (CGN) ↔ Lys (AAR)) from optimal. A polarity/hyrophobicity cluster formed by NAN, NGN and NUN codons was weaker, requiring exchange of 2 pairs to reach an optimal distribution. In relation to the former, there are 2.52 x $10^3$, or 7!/2!, possible codes. There are 24, or $(2)^2$ x 3!, equally clustered codes, and 12, or 2 x 3!, more clustered than the stage 14 code for ANN, CNN and UNN sets, apparent on inspection of Fig. 1. The probability of this polarity/hydrophobicity gradient forming by chance is 1.43 x $10^{-3}$ (Table 1).

Combining probabilities obtained at each stage of code evolution (Table 1), by the method of Fisher (1959), revealed biosynthetic relations were highly significant (P = 8.96 x $10^{-6}$) during evolution of the genetic code. Inclusion of acceptor core groups (Saks & Sampson, 1995) elevated the significance of correlations between biosynthetically related amino acids (P = 4.42 x $10^{-8}$), by more than two orders of magnitude. This adds credence to the proposition that amino acid acceptors served as cofactors in early amino acid synthesis. Charge confinement was optimal in the N-fixers code, and the effect approached statistical significance on combining stage 2 and 14 probabilities (P = 0.106). This fits with the significance attributed to charge attraction between a polypeptide chain and positively charged mineral surface in the early stages of code evolution. As deduced from the standard code by previous investigators, residue polarity/hydrophobicity was found to have played a highly significant role in the evolution of the genetic code (P = 8.98 x $10^{-5}$). This was most noticeable during code expansion (stage 4 to 7), when formation of hydrophobic domains in emerging globular and membrane proteins appears to have directed evolution of the genetic code.

## 6. **Discussion**

The biosynthetic model of code evolution reveals why codons of some precursor/product amino acid pairs in the standard code match, while others do not. Broadly, a change in the direction code evolution and in the mechanism of codon assignment determined the outcome. Expansion from the N-fixers code, between stage 4 and 7, was found to proceed by codon assignments along the rows GNN, ANN, CNN and ANN. Earlier, codon assignments were limited to a single column (NAN). A change in direction of code evolution resulted, It led to Pro$^{4}$ and Thr$^{6}$ being assigned codons (CCN, CAN) that differed at two sites from codons (GAR, GAY) for their primary precursor, Glu$^{1}$ and Asp$^{1}$, respectively.

Post-expansion (stage 9 to 14) additions to the code required codon reassignment. When an incoming amino acid captured an unrelated acceptor, any agreement between precursor and product codons would be purely random. Post-expansion entrants to the code included Arg$^{9}$ and Lys$^{10}$ (Davis 1999; Fig. 1). They have codons, CGN, AGR and AAR, with two, and three, base differences from codons for their primary precursor, Glu$^{1}$ and Asp$^{1}$, respectively. Wong (1975) omitted Glu/Pro, Glu/Arg, Asp/Thr and Asp/Lys, together with Thr/Met, when seeking to match base triplets among precursor/product pairs. Their inclusion appeared to necessitate "plausible, but less certain" additions to Glu and Asp codons. It is now apparent, however, that the biosynthetic model provides justification for their exclusion, without an ad hoc expansion of the codon sets for Glu and Asp.

Effects of a change in direction of code growth and mechanism of codon assignment were avoided by considering each phase identified in code evolution independently. tRNA core structure also provided a guide to acceptor origin and selection of antecedent/amino acid pairs. Highly significant correlations resulted between base triplets assigned to all biosynthetically related amino acids (Table 1). The combined phase-dependent probability for codon homology and codon + tRNA core group homology among biosynthetically related amino acids was $P = 8.96 \times 10^{-6}$ and $P = 4.43 \times 10^{-8}$, respectively. This is about three and five orders of magnitude lower than the 'before and after' probability in 5'- (P = 9.8 x $10^{-3}$) and mid-base (P = 7.8 x $10^{-3}$) in codons of 11 precursor/product amino acid pairs (Davis, 1999, §5). Non-significantcodon correlations among an inclusive set of precursor/product pairs in the standard code (Amirnovin, 1997; Ronnenberg et al., 2000) can now be attributed to large, unaccounted sources of variation. Di Giulio & Medugno noted that the probability for observed codon homology decreased from 0.12 (Fig. 2) to 6 x $10^{-6}$ (Di Giulio & Medugno, 2000, Table 1), on passing from code-wide to combined family-wide correlations within the standard code.

The early investigation by Wong (1975) contained two unmet preconditions: (i) Identification of the correct antecedent for establishing codon correlations, and (ii) removal of large non-biosynthetic sources of codon variability. In a sense, Wong chose a parameter that was too sensitive for the information available to him in 1975. Significant probabilities characterize the 'before and after' order of codons for precursor/product pairs (Davis, 1999). Had he chosen to base the biosynthetic model on this parameter, the chemo-autotrophic model for the origin of life is likely to have been elucidated a decade before pathway retrodiction led to it (Wachterhauser, 1988, 1992, 1997, 1998).

Codons of the NAN set occurred as the earliest part of the code, in an analysis of codons for precursor/product amino acid pairs (Davis, 1999, §5). They encode N-fixing amino acids, and, consistent with their antiquity, the N-fixers include precursors, in contemporary pathways, to nearly half the amino acids in the code. Furthermore, they are synthesized on short pathways, within two reaction steps of the citrate cycle. This cycle autocatalytically fixes $CO_2$, under reducing, early Earth conditions. Carbonic anhydrase activity intrinsic to these amino acids (Bar-Nun et al., 1994) could thus accelerate the reductive citrate cycle and, thereby, boost their own synthesis. The genetic code had thus preserved evidence of the origin of life and, in doing so, validated the chemo-autotrophic model for the origin of life (Wachterhauser, 1988).

There is now convincing evidence that amino acid acceptors served as cofactors in amino acid synthesis throughout code evolution. Even in the small N-fixers code, where codon homology among biosynthetically related amino acids was obscured (Table 1), biochemical observations establish that tRNA serves as a cofactor in prokaryote synthesis of both Asn and Gln (Wilcox & Nirenberg, 1968; Schon et al., 1988; Curnow et al., 1989, Gagnon et al., 1996). Participation of tRNA cofactors in early amino acid synthesis is consistent with a free carboxyl over the length of extant pathways for amino acid synthesis, excluding His[13]. The 'protected' carboxyl in these pathways may then viewed as a relic from RNA cataylzed synthesis, when it served as an attachment site for tRNA cofactors. By contrast, formation of the $\alpha$-amine occurs at, or near, the last step in the synthesis of most amino acids; the Asp[1] $\rightarrow$ Thr[6] path being the only notable exception. From the present perspective, this appears to represent a mechanism to prevent intermediates, attached to a tRNA cofactor, from being incorporated into protein

FIGURE LEGEND

Fig. 1. Main stages identified in the evolution of the genetic code. Pre-code synthesis is shown to lead to polypeptide chains containing the four N-fixing amino acids in random sequence. AAA triplets in a poly A template were evidently read by a single tRNA species with a group D core, after being charged with any one of these amino acids. Increased efficiency of N-fixation in a primordial system on a mineral surface accompanies polycondensation of Asp, Glu, Asn and Gln in an ordered sequence. Triplets in the NAN set were found to be assigned to each of the four amino acids and Ter in the N-fixers code. The three acceptors for Asp, Glu and Gln had a group D core, as in ancestral tRNA. The Asn acceptor had group A core. As it shared anticodon UUU with ancestral tRNA, its unique core structure provided a recognition element for specific amino acylation. The N-fixing amino acids are synthesized on paths extending 1 or 2 reaction steps from the autocatalytic reductive citrate cycle. Expansion of the code occurred through assignment of NCN, NGN and NUN triplets to increasingly hydrophobic (more negative transfer free energy, $\Delta F_T$) amino acids. Synthesis of these aliphatic amino acids originated at branch reactions in central metabolism on paths extending 4 to 7 steps, measured from the citrate cycle. Acceptors with a group C core appeared in this stage. Serine$^4$ and Leu$^7$ were loaded onto a modified acceptor (type II) and each read two sets of codons. In the post-expansion phase stage, amino acids with side chains containing basic groups and ring structures were incorporated into protein. They entered the code by capturing assigned codons. Their formation occurred on paths of 9 to 14 steps, measured from the citrate cycle. Acceptors with a group B core appeared at this stage. Superscripts refer to the number of steps, from the citrate cycle, in amino acid synthesis, or, equivalently, to the stage of code evolution . Arrows connect biosynthetically related amino acids whose acceptors contain the same core group.

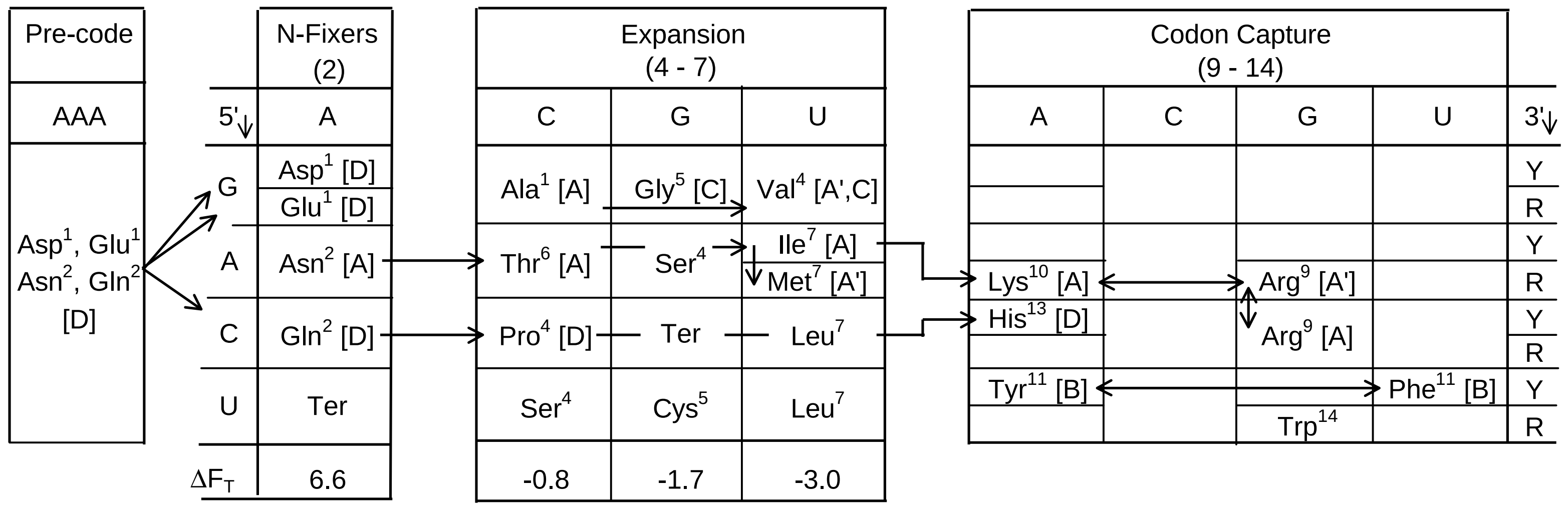

Table 1. Biosynthetic and physical chemical patterns formed by codon assignments at three stages during evolution of the genetic code

| Code | Stage | Cluster[a] size n | Cluster[a] extent $X_o \pm C$ | No. of possible codes | No. codes with $X \leq X_o$ | P[c] | Cluster[a] size n | Cluster[a] extent $X_o \pm C$ | No. of possible codes | No. codes with $X \leq X_o$ | P[c] |
|---|---|---|---|---|---|---|---|---|---|---|---|
| | | *Biosynthetic kinship* | | | | | *Biosynthetic kinship with related tRNA* | | | | |
| N-fixers | 2 | 2 | 0.5±1.41 | 480 | 48 | 0.10 | 2 | 1.0 | $7.68\times10^{3}$ | 576 | $7.50\times10^{-2}$ |
| Expansion | 7 | 10 | 1.3±0.35 | $2.39\times10^{9}$ | $6.91\times10^{5}$ | $2.89\times10^{-4}$ | 10 | 1.7± 0.40 | $6.03\times10^{17}$ | $1.63\times10^{12}$ | $2.71\times10^{-6}$ |
| Capture | 14 | 6 | 1.4±0.64 | $4.65\times10^{10}$ | $1.31\times10^{7}$ | $2.91\times10^{-4}$ | 6 | 1.6±0.56 | $7.44\times10^{15}$ | $2.61\times10^{8}$ | $3.51\times10^{-8}$ |
| | | $\chi^2_6 = 37.19^{***}$, $P = 1.62\times10^{-6}$ | | | | | $\chi^2_6 = 65.15^{***}$, $P = 4.02\times10^{-12}$ | | | | |
| | | *Charge confinement* | | | | | *Polarity/hydrophobicity*[b] | | | | |
| N-fixers | 2 | 2 | 1.0 | 120 | 12 | 0.10 | 4 | 1.0 | 120 | 36 | 0.30 |
| Expansion | 7 | | | | | | 10 | 1.2± 0.68 | $1.20\times10^{8}$ | $3.89\times10^{4}$ | $3.25\times10^{-4}$ |
| Capture | 14 | 4 | 1.5±0.50 | $6.0\times10^{3}$ | 804 | 0.15 | 6 | 1.5± 0.38 | $2.52\times10^{3}$ | 36 | $1.43\times10^{-2}$ |
| | | $\chi^2_4 = 7.62$ NS, $P = 0.11$ | | | | | $\chi^2_6 = 28.10^{***}$, $P = 8.98\times10^{-5}$ | | | | |

[a] Cluster size, n, is the number of amino acids (Polarity/hydrophobicity), pairs of related amino acids (Biosynthetic kinship, Biosynthetic kinship with related tRNA), and sets of codons assigned to charged amino acids (Charge confinement). Cluster extent refers to the mean ± coefficient of variation of differences in codon bases (Biosynthetic kinship), codon bases plus acceptor core groups (Biosynthetic kinship with related tRNA), bases of codon sets (Charge confinement), and increments between residue rank on a transfer free energy scale in specified codon sets (Polarity/hydrophobicity). $X_o$ is the observed mean difference.

[b] Polarity/hydrophobicity gives the probability of a code exhibiting an equal, or smaller, distance (increment in free energy rank) among residues assigned to specified codon sets. These sets were (GAN, GAY) and (AAN, CAN) in the N-fixers code, NCN, NGN and NUN in expansion phase, and ANN, CNN, GNN and UNN during codon capture. The free energy changes refer to transfer of each residue from an aqueous to non-polar (dielectric constant, 2.0) medium (Tolstrup et al., 1994).

[c] P is the probability of a code of specified extent, $X < X_o$. Procedures for calculating the number of possible codes and number of equal, or more, clustered codes are given in the text. $\chi^2$ values were obtained using Fisher's method for combining probabilities at each phase of code evolution, with an indicated number of degrees of freedom (Fisher, 1959). NS implies $P > 0.05$, and $^{***}$, $0.01 > P$.